\documentclass[reprint,aip,amsmath,amssymb,color,longbibliography,superscriptaddress]{revtex4-2}
\usepackage{stmaryrd}
\usepackage{amsmath}
\usepackage{amssymb}
\usepackage{graphicx}
\usepackage{dcolumn}
\usepackage{bm}
\usepackage{tabularx}
\usepackage{color}

\begin{document}
\begin{titlepage}
\title{Bond Dipole based Geometric Theory of Band Alignment}
\author{Zeyu Jiang}
\affiliation{Department of Physics, Applied Physics and Astronomy, Rensselaer Polytechnic Institute, Troy, NY, 12180, USA}
\author{Damien West}
\affiliation{Department of Physics, Applied Physics and Astronomy, Rensselaer Polytechnic Institute, Troy, NY, 12180, USA}
\author{Shengbai Zhang}
\email{zhangs9@rpi.edu}
\affiliation{Department of Physics, Applied Physics and Astronomy, Rensselaer Polytechnic Institute, Troy, NY, 12180, USA}
\date{\today}

\begin{abstract}

The band alignment (BA) between two materials is a fundamental property that governs the functionality and performance of electronic, as well as electrochemical, devices. However, despite decades of study, the inability to separate surface properties from those of bulk have made a deep understanding of the physics of BA illusive. Building on the theory of ideal vacuum level to separate surface from bulk [CWZ, Phys. Rev. B {103}, 235202 (2021)], here we present a geometric theory for the band alignment, particularly, explaining the insensitivity of the alignment to interfacial orientation between isotropic materials. First, we adopt charge neutral polyhedron, termed Wigner-Seitz atoms (WSA), to partition the charge of atoms in a way which maintains crystal symmetry and tessellates the space. In contrast to CWZ theory, the band alignment of two materials constructed from such WSAs is independent of interface orientation. Upon electron relaxation at the interface, here we show that the interfacial charge transfer dipole can be faithfully descibed by the sum of localized point dipoles which exist between atoms at the interface (bond dipoles). For interfaces between isotropic materials, the magnitude of the bond dipole can be factored out as a multiplier, leaving only geometric factors, such as the crystal symmetry and dimension of the material, to determine band alignment, irrespective of the orientation of the interface. We considered 29 distinct interfaces and found that this bond dipole theory yields excellent  agreement (RMS deviation $<$ 30 meV) with first-principles results. Our theory can be straightforwardly applied to interface between alloys, as well as between anisotropic systems.

\end{abstract}

\maketitle
\draft
\vspace{2mm}
\end{titlepage}

\section{Introduction}
With the rise of micro and quantum electronics, understanding the interface at the atomic level can be critically important for the semiconductor industry\cite{ohno1999, chakhalian2012, hwang2012, alferov1998, wang2014}. Of key importance to interface electronics is the band alignment which determines the performance threshold of quantum devices\cite{kamat2012,chuang2014,shao2020,li2021,sun2021}. Since the 1980s, a wide number of experiments\cite{wang1985,langer1985,hirakawa1990,becker1991,yu1992} have revealed that the band offset of conventional semiconductors are largely independent of interface orientation. These results are quite surprising given that different orientations are associated with different concentrations of interfacial bonds and hence the interfacial charge transfer (and the resulting interfacial dipole) contribution to the band offset is expected to depend strongly on the interfacial details. How the bulk and interfacial properties of heterostructures conspire to yield nearly orientation independent band alignment remains a deep mystery.

On the theoretical side, first-principles methods like density functional theory (DFT) can provide fully quantum mechanical treatment of the band offset of specified materials using a periodic supercell containing an atomically constructed interface\cite{walle1987,baldereschi1988,christensen1988a,christensen1988b,wei1998,Li2009}. These calculations provide high quality results and reproduce the experimentally observed orientation independence. However, as such approaches merely output the self-consistent ground state electronic configuration of the interface, they provide very limited insight into the underlying physics of interface formation. Nevertheless, from such calculations, it is clear that the dipole formation which governs the band alignment is quite local in nature (typically on the order of a bond length), in contrast to the interfacial state wavefunction which is much more delocalized across the interface\cite{zhang1990,zhang2007}. 

One of the most famous examples to understand band alignment is the Anderson's rule\cite{anderson1960} which proposes that two isolated semiconductors are aligned through their workfunctions. Typically, Anderson's rule works well for van der Waals-coupled low dimensional systems\cite{besse2021} but fails in covalent systems as the assumption is essentially that no charge rearrangement takes place when the two systems are brought together, which would lead to a dipole at the interface. Further, this highlights the fundamental issue that the local interfacial relaxation dipole can only be defined with respect to a \emph{reference} configuration, and it is unclear what reference system can provide the most physical insight.

Quite another approach was proposed by Tersoff in Ref.\cite{tersoff1984a,tersoff1984b}, wherein the band alignment is entirely a property of the individual bulk materials. Instead of determining charge transfer at the interface, the bulk band structures of two semiconductors are said to be aligned at their respective charge neutrality levels. While this branch point theory is widely cited to understand orientation independence of the interface properties, there exist multiple conflicting definitions of the charge neutrality level in the literature\cite{tersoff1985,schleife2009,Hinuma2014} and its validity is difficult to experimentally access. A perspicuous theory of interface physics with explicit separation between bulk and interfacial contributions, at even the $\it{semiclassical}$ level, is largely lacking.

To remedy this, considerable efforts\cite{cardona1987,jaros1988,lambrecht1990,walle1987} have been made to separate the bulk contribution and clarify the mechanism of interfacial dipole relaxation. However, while such studies agree on the band offset, different choices of reference lead to strongly conflicting values for the contribution of interfacial dipole. While some methods, such as superposition of atomic charge, successfully remove orientation dependence of the bulk contribution to band offset, they do not accurately represent the bulk charge densities of the constituent materials and are hence difficult to justify. More recently, reference systems have been developed which faithfully describe the charge density in the bulk region. Constructing an interface from a planar cut at the maximum of the planar-average potential\cite{choe2018,choe2021} has been used to define an ideal vacuum level, wherein it was shown that the band offset of several semiconductor interfaces are well described by the associated ideal vacuum alignment. However, as the ideal vacuum level was found to depend strongly on orientation, the cause of the cancellation of this orientation dependence in the band offset is unclear. 

Instead of using a planar cut, if one first partitions the bulk charge density into regions that each atom can be associated with a volume, the reference system can then be constructed as an interface built from these atomic volumes. The most well-known way to associate bulk charge with a particular atom is from Bader\cite{bader1990}, however, as each atomic volume is charged, the unit cell has a dipole moment and hence the bulk average potential and associated ideal vacuum level is ill-defined\cite{choe2018,choe2021}. An alternate partitioning was proposed by Tung and Kronik \cite{tung2014,tung2016} wherein the charge density is split into atom-centered charge neutral polyhedra (NP), with the advantage that the NP band offset, before electronic (not to mention ionic) relaxation, is independent of orientation. In examining a range of zincblende \cite{tung2016} and perovskite \cite{tung2019} hetrostructures, they found that the band offsets of fully relaxed heterostructures were largely independent of orientation and fairly well described by the NP band offsets. They attributed \cite{tung2019} the \emph{approximate} orientation independence of band offsets of fully relaxed interfaces to the ``ionic screening'' effect wherein the charge transfer dipole which develops due to electron relaxation (electron transfer dipole) is largely cancelled by the charge transfer dipole due to ionic relaxation. As a result, the band offset is primarly determined by the NP band offset.

In this work, we find that the electron transfer dipole which develops at the interface between two isotropic materials will \emph{itself} be independent of orientation, indicating that for common semiconductors this orientation independence arises from the underlying crystal symmetry instead of from ionic screening. Using density functional theory (DFT) with a NP bulk charge reference, we examine 29 distinct interfaces between systems including diamond, zincblende, rock-salt, and cesium chloride lattices in three dimensions (3D),  square and hexagonal lattices in two dimensions (2D), and stacked 2D systems. We show that the interfacial dipole which develops due to electron relaxation can be understood entirely in terms of the sum of \emph{point bond dipoles} which form along the individual interfacial bonds. A general geometric model for the interfacial dipole, in terms of the bond dipoles, is constructed and allows for determination of the band alignment for an interface of arbitrary orientation. Analytic solutions to this model correctly predict the orientation dependence of the band alignment in anisotropic materials and shows that the orientation independent band alignment of isotropic materials arises from the underlying crystal symmetry. In contrast, analysis of subsequent ionic relaxation tends to be orientation dependent even for isotropic materials,  leading to a certain degree of orientation dependence of the band alignment, as seems to be the case in  experiment.

This paper is organized as follows: Sections (A) and (B) describe the construction of Wigner-Seitz atoms and solids and describe the band alignment in the absence of any interfacial charge transfer. Sections (C-E) focus on presenting the bond dipole theory in which interfacial charge transfer is described in terms of local point dipoles between atoms. Section (F) discusses the extension of the above theories to isotropic alloys and beyond. Section (G) discusses the contribution of ionic relaxation to band alignment.

\section{Results and Discussion}

\subsection{Wigner-Seitz atom and solid}

\begin{figure}[tbp]
	\includegraphics[width=\columnwidth]{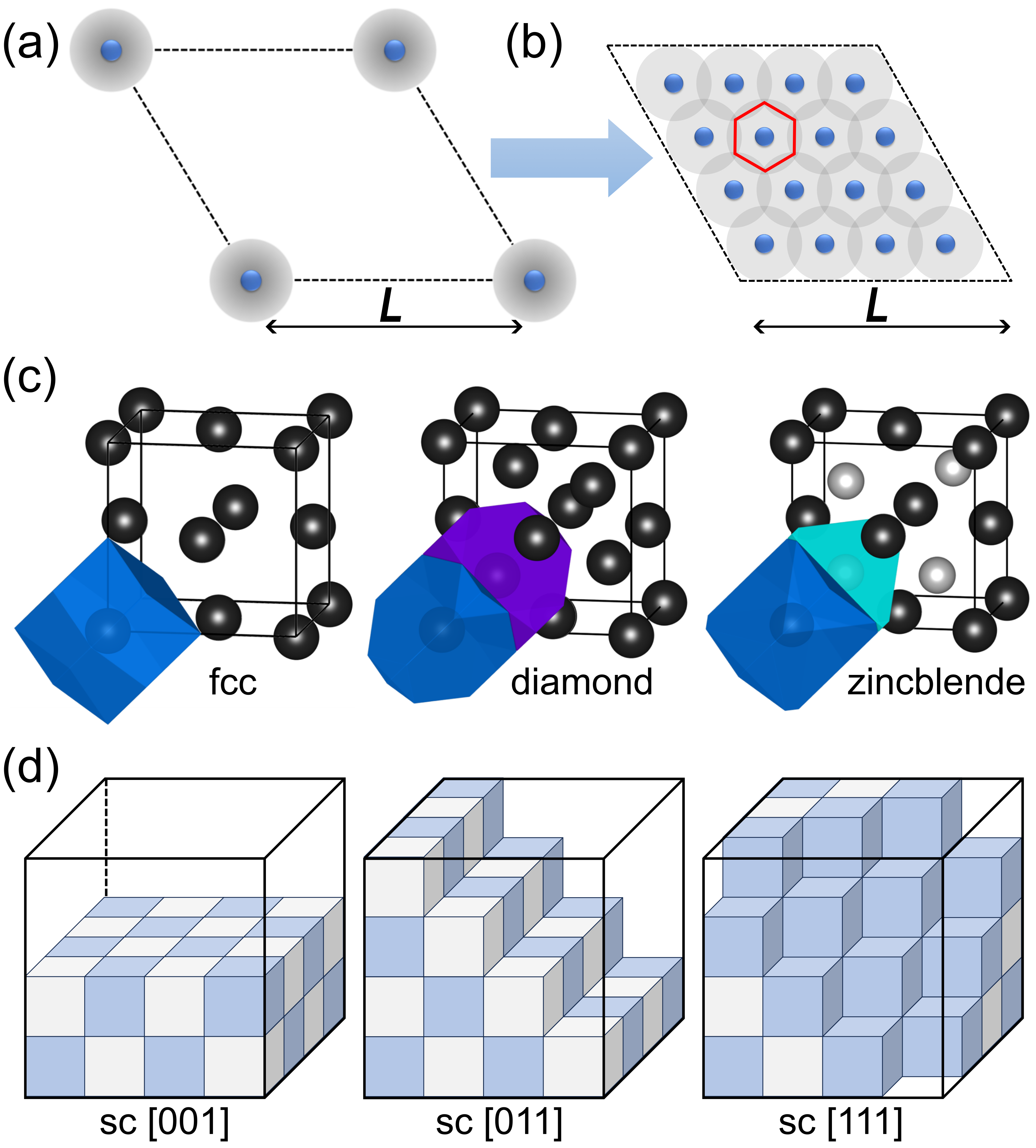}
	\caption{\label{fig:fig1} \textbf{Partition of a solid with Wigner-Seitz atoms.} (a) Schematic of far separated atoms with spherical charge distribution. (b) Schematic of solid formation with continuously distributed charge density, induced by the overlap and relaxation of atomic charge under the bulk crystal symmetry. (c) Wigner-Seitz atoms of face-centered cubic (fcc), diamond and zincblende lattices. (d) Surfaces formation by assembling Wigner-Seitz atoms of simple cubic lattice in [001], [011] and [111] directions. Alternating colors are used to distinguish between the nearest neighboring Wigner-Seitz atoms.}
\end{figure}

The formation of an interface has two steps: (1)A surface is truncated from periodic bulk with fixed charge density and the two truncated surfaces are merged into an interface, (2)Interface dipoles are formed when the truncated charge density at the interfaces relaxes to equilibrium. The former determines the bulk contribution to band offset while the latter is the interfacial effect which provides an additional potential shift across the interface. This has motivated the introduction of ideal vacuum level, which is determined as the maximum of planar averaged potential $\overline{V}_{\hat{\bm{n}}}$ for any direction $\hat{\bm{n}}$. Here the bulk electron density is truncated by a plane parallel to the interface at the position of $\overline{V}_{\hat{\bm{n}}}$ maximum and then re-connected to each other at the cutting plane. However, as oblique angles can lead to cutting planes which cut close to atomic core regions, this leads to high energy reference systems which seem quite unphysical resulting in highly direction dependent bulk contribution to the band offset (as detailed in Sec. II. B). As a result, the electron relaxation is very large and predominately associated with regaining the approximate spherical atomic symmetry which is broken by the planar cut. As such, a charge partitioning method which avoids macroscopic fields and respects both the translational crystal symmetry as well as the local symmetry of atoms is most desirable as it is expected to minimize both the magnitude and orientation dependence of subsequent electron relaxation, allowing for the study of charge transfer along the bond, as detailed in the next section.

For atoms which are far separated, as in Fig. 1 (a), the charge density can be considered spherical and directly associated with a particular ion, however, as the spacing is decreased to form a solid, as in Fig. 1 (b), the charge densities of adjacent atoms overlap, lowering the local atomic symmetry to that of the crystal as shown by the red hexagon. Due this overlap, restoring spherical symmetry of the atomic charge is neither well defined, nor can be accomplished through real-space charge truncation. Instead, reminiscent of the concept of Wigner-Seitz unit cell\cite{wigner1933} which reassigns the volumes of crystal into lattice-point centered polyhedrons, we partition the atomic charge to atom-centered polyhedra which respect the crystal symmetry. 

Wigner-Seitz unit cell is defined for Bravais lattices, however, for a composite lattice which has multiple atoms in each unit cell, the volume of Wigner-Seitz unit cell has to be further decomposed into even smaller polyhedrons centered on each atom\cite{grosso2014}. Such atom-specific polyhedron, here dubbed as Wigner-Seitz atom, refer to the volume contained within the cutting planes which are perpendicular to the lines connecting neighboring atoms in which the cutting plane distance along the line is determined by the requirement of charge neutrality. The volume of Wigner-Seitz atom depends not only on the lattice structure but also the relative electron affinities of atomic elements, since each Wigner-Seitz atom has to be charge neutral individually. As exemplified in Fig. 1(c), the Wigner-Seitz unit cell of face-centered cubic (fcc) lattice is a rhombic dodecahedron, while the Wigner-Seitz atom of diamond structure is a triakis truncated tetrahedral honeycomb\cite{conway2008,tung2014}. More generally, the volume of Wigner-Seitz atom will be element-specific for a compound due to the varied electron affinities of different species and the required charge neutrality of each individual Wigner-Seitz atom, as shown by the zincblende structure in Fig. 1(c). While we prefer the name Wigner-Seitz atom as it makes clear that lattice symmetry is respected, such charge partitioning has been referred to as neutral polyhedra by Tung et al.\cite{tung2016,tung2019}. By a tessellation of Wigner-Seitz atoms in real space, the bulk charge density can be reproduced without any overlap or void, leading to a Wigner-Seitz solid. In such a solid, the charge density associated with each ion is tightly packed around it, leading to a systematically smaller quadrupole than that of an isolated atom of the same specie [Note S1 in the supplementary materials (SM)].

\subsection{Wigner-Seitz interface}

\begin{figure}[tbp]
	\includegraphics[width=\columnwidth]{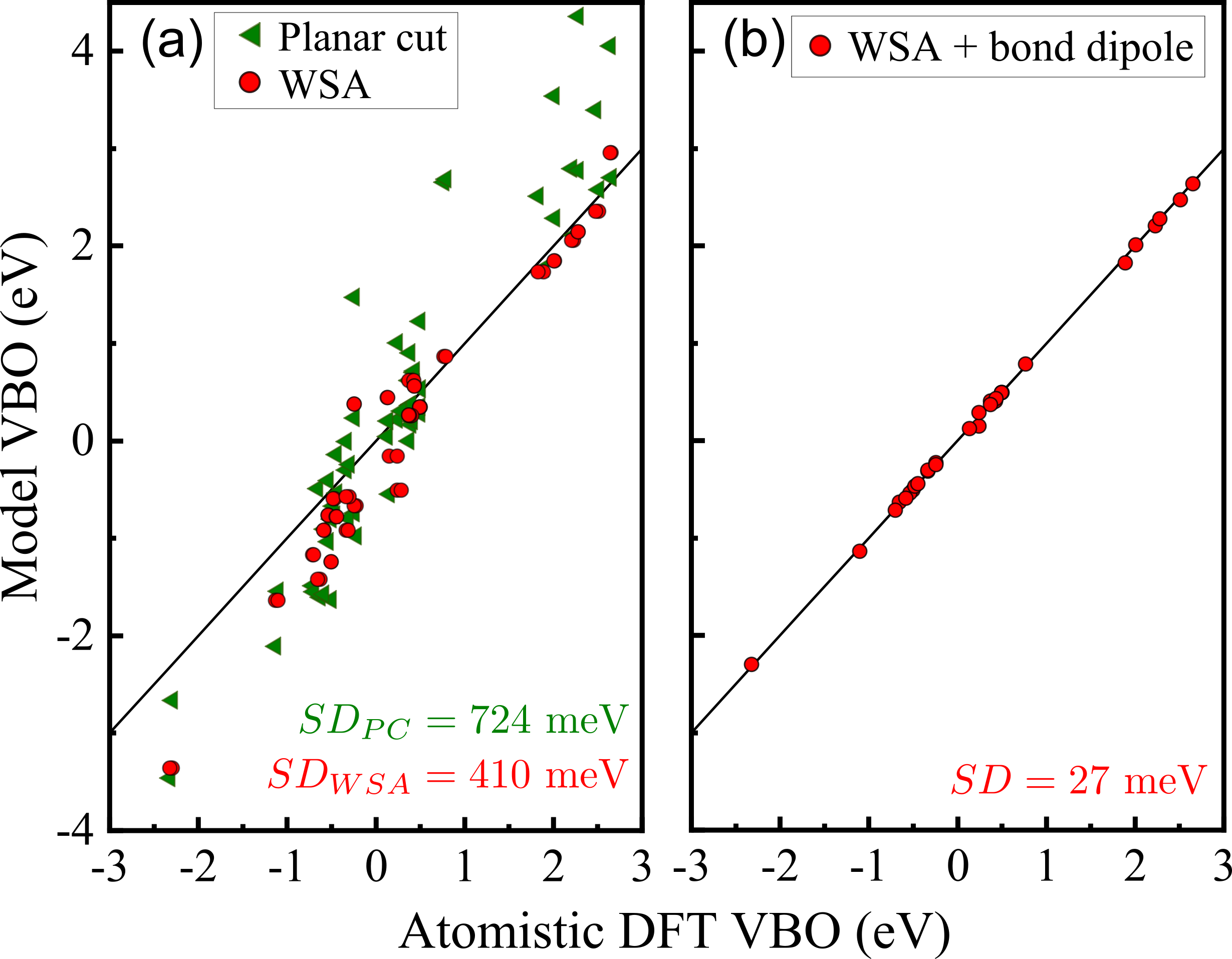}
	\caption{\label{fig:fig2} \textbf{Valence band offset (VBO) calculated for 28 different semiconductor interfaces.} Horizontal axis denotes the results of atomistic DFT supercell calculations. (a) Vertical axis denotes the results obtained by connecting two truncated surfaces without any interfacial electronic and ionic relaxation. Red circles are the results of WSA interfaces. Green triangles are the results of planar cut (PC). (b) Vertical axis denotes the results that the potential shift due to the present of bond dipoles (BD) is added to WSA results (WSA+BD). Since more than one directions are considered for each interface system, the total number of data points is larger than 28. Details of these interface systems and their VBOs are summarized in Table S1 of the SM.}
\end{figure}

A Wigner-Seitz surface can be defined for any arbitrary direction by partitioning the bulk charge density along the boundaries of Wigner-Seitz atoms, as exemplified for a simple cubic lattice in Fig. 1(d), and a Wigner-Seitz interface is constructed by aligning up two such surfaces without relaxing the interfacial charge. Here we show that the Wigner-Seitz interface is an excellent reference to separate bulk and interfacial contributions to interface formation, since such developed interfacial dipoles are small and more importantly for isotropic systems, it leads to rigorous orientation independence of the band alignment.

This can be shown by considering the band alignment of the Wigner-Seitz interface before electron relaxation. To determine the band offset, we first consider the average potential $\overline{V}_{bulk}$ of each material relative to the associated vacuum ${V}_{vac}^{\hat{{n}}}$ in the vicinity of the surface (with surface normal $\hat{{n}}$). Here we note this quantity ${V}_{vac}^{\hat{{n}}} - \overline{V}_{bulk}$ is typically strongly dependent on surface orientation \cite{choe2018,choe2021,jiang2023} and is given by

\begin{equation}
	\begin{aligned}
{V}_{vac}^{\hat{{n}}} - \overline{V}_{bulk} &= \frac{1}{2 \epsilon_0 \Omega} \hat{{n}}^{T} \overleftrightarrow{{Q}} \hat{{n}} \\
    &= \frac{1}{2 \epsilon_0 \Omega}  \sum_{i} \int_\Omega d^{3}{r} \left(\hat{{n}} \cdot {\vec{r}}\right)^2 \rho_{i}\left({\vec{r}}\right),
    \end{aligned}
\end{equation}
where $\Omega$ is the unit cell volume, $\overleftrightarrow{{Q}}$ is the \emph{tracefull} electric quadrupole with $Q_{lm}=\int_\Omega r_l r_m \rho({\vec{r}})d^{3}{r}$, and $\rho_{i}\left({\vec{r}}\right)$ is the charge density of Wigner-Seitz atom of $i$-th species (e.g., $i=\{1,2\}$ for binaries). The potential shift ${V}_{vac}^{\hat{{n}}} - \overline{V}_{bulk}$ is purely determined by the bulk properties and dominates the bulk effect of interface formation. We note that the potential shift indicated by Eq. (1) is the \emph{electrostatic potential}, when given in units of electron volts (eV) it has been multiplied by $-e$ to represent the electron energy.

By expanding Eq. (1) we have
\begin{equation}
	\begin{aligned}
	{V}_{vac}^{\hat{{n}}} - \overline{V}_{bulk} = &\frac{1}{2 \epsilon_0 \Omega} \sum_{i} \int_\Omega d^{3}{r} (\alpha^2 x^2 + \beta^2 y^2 + \gamma^2 z^2)\rho_{i}\left({\vec{r}}\right) +\\ & \frac{1}{\epsilon_0 \Omega} \sum_{i} \int_\Omega d^{3}{r}(\alpha\beta xy + \beta\gamma yz + \alpha\gamma xz) \rho_{i}\left({\vec{r}}\right),
    \end{aligned}
\end{equation}
where $\alpha$, $\beta$ and $\gamma$ are the $x$, $y$ and $z$ components of normal vector $\hat{{n}}$ with $\alpha^2+\beta^2+\gamma^2=1$. The first and second terms on right correspond to the diagonal and off-diagonal elements of the quadrupole tensor. For isotropic systems with equivalent $x$, $y$ and $z$ axes,
\begin{equation}
\int_\Omega d^{3}{r} x^2\rho_{i}\left({\vec{r}}\right) = \int_\Omega d^{3}{r} y^2\rho_{i}\left({\vec{r}}\right) = \int _\Omega d^{3}{r} z^2\rho_{i}\left({\vec{r}}\right) =Q_i,
\end{equation}   
where $Q_i$ is the bulk electric quadrupole of $i$-th species. Taking into account the two-fold rotational symmetry along three axes of an isotropic system, the off-diagonal terms in Eq. (2) will vanish. Finally, for isotropic systems Eq. (1) turns out to be
\begin{equation}
	{V}_{vac}^{\hat{{n}}} - \overline{V}_{bulk} = \frac{1}{2 \epsilon_0 \Omega} \sum_{i} Q_i,
\end{equation}
which is rigorously orientation independent. By constructing a Wigner-Seitz interface, the vacuum levels of two Wigner-Seitz surfaces align with each other, leading to an orientation independent band offset between two materials for isotropic systems. 

Figure 2(a) shows the valence band offsets (VBO) obtained from Wigner-Seitz interface and planar cut alignment compared to atomistic DFT supercell calculations for 28 distinct interfaces of materials with a wide range of structures and properties, including diamond, zincblende, rock-salt, and cesium chloride lattices in 3D and square and hexagonal lattices in 2D. Here we can see that in addition to being independent of orientation, the VBO predicted by WSA has less deviation from the atomistic DFT results (closer to the line) than the planar cut, having a root-mean-square deviation of 410 meV vs. 724 meV. As such, the interfacial dipole associated with subsequent electron relaxation is smaller for the WSA reference than the planar cut. 

While WSA seems to minimize the subsequent electron relaxation, we note that the interfacial dipole arising from the electron relaxation can still shift the VBO by more than 1 eV. Nevertheless, by choosing this Wigner-Seitz solid reference, we see that not only does the bulk contribution to the VBO become orientation independent but that a significant portion of the VBO is already captured by these bulk properties. As such, the subsequent electron relaxation is associated with local chemistry at the interface enabling for the formulation of our bond dipole theory of electron relaxation at the interface, to be descibed in the subsequent sections.

\subsection{Interfacial bond dipole}

\begin{figure}[tbp]
	\includegraphics[width=\columnwidth]{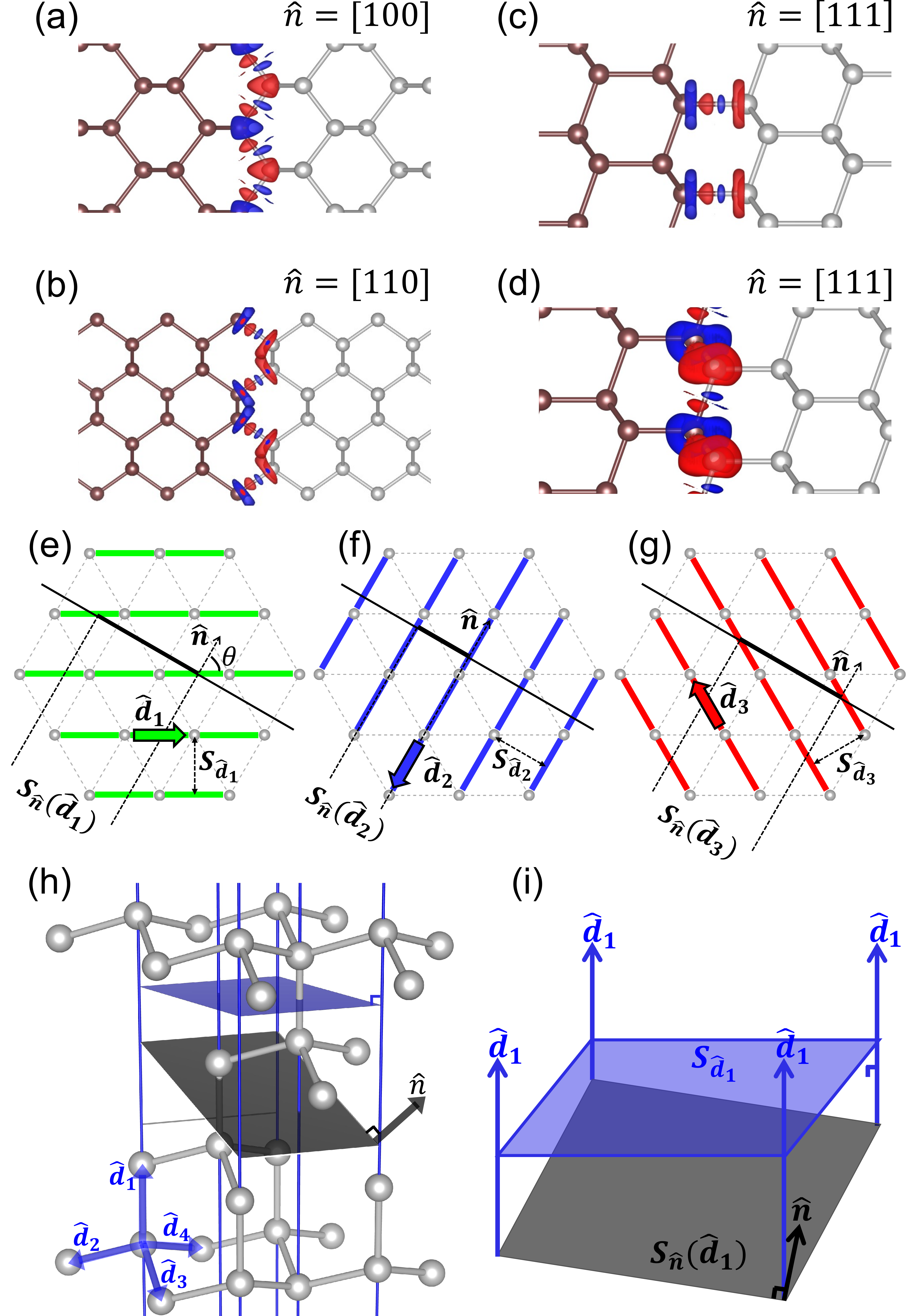}
	\caption{\label{fig:fig3} \textbf{Interfacial bond dipole in the monoatomic system.} (a-d) The charge distribution of bond dipoles at Si(left)-Ge(right) interface along (a) [100], (b) [110], and (c-d) [111] direction. Red (blue) cloud denotes the positive (negative) dipole charges. (e-g) Schematic of bond lines cutting by an interface line in 2D triangular lattice. The green, blue and red arrows denote three different bond directions, respectively. (h) Schematic of chemical bonds truncation in 3D diamond lattice. (i) An abstracted illustration of (h) to show the relationship between $S_{\hat{{n}}}(\hat{{d}})$ and $S_{\hat{{d}}}$.}
\end{figure}
Even when considering the interface between two materials which are lattice matched, the details of the geometry and atomic coordination at the interface depends on its orientation. While the band alignment of Wigner-Seitz interface can be rigorously isotropic, the onset of interfacial dipole relaxation could in principle break the symmetry and lead to anisotropic results. 

Using the diamond structure Si-Ge as an example of a protypical interface, the charge density associated with the interfacial dipole is depicted in Figs. 3(a-d). The charge density of interfacial dipole is defined as $\Delta \rho = \rho[AB]-\rho[A]-\rho[B]$ for the A-B interface, where $\rho[AB]$ is the charge density of the interface with electronic density being fully relaxed and $\rho[A]$ ($\rho[B]$) is the charge density of Wigner-Seitz surface assembled by Wigner-Seitz atoms of material A (B). Due to charge neutrality $\Delta \rho$ does not contain a net charge, but only electric dipoles caused by interfacial electron relaxation. Here, we have used a lattice matched Si-Ge interface constructed with the average lattice constant. 

We find that the charge relaxation at the interface occurs along the bonds.  In such a monoatomic interface, an alternate charge neutral partitioning (Bader volumes) is also possible in which the cutting surfaces are not planar. However, we find that the electron relaxation using such a Bader reference is substantially more delocalized in regions where the flat surface has been replaced by the curved surface, see Note S2 in the SM. As such, this suggests that the WSA reference is a good choice to study the interface, motivating us to postulate that when using a WSA reference, the effect of the electron relaxation can be modeled as a collection of point bond dipoles, $\vec{{p}}$, which are highly localized on each bond at the interface and pointing along the bond direction $\hat{{d}}$. 
It is important to distinguish between the bond dipole $\vec{{p}}$ along the interfacial bond and the net interfacial dipole perpendicular to the interface which directly contributes to the band offset. The magnitude of the bond dipole is driven by the overlap (or void) between Wigner-Seitz atoms of different species across the interface. However, the contribution of this bond dipole to the band offset is determined by its projection in the direction normal to the interface. From a simple capacitor model, the potential shift due to $\vec{{p}}$ across the interface is
\begin{equation}
	V_{D_{R}}^{\hat{{n}}}=\frac{\sigma_{\hat{{n}}}}{\varepsilon_{0}}, 
\end{equation}
where $\varepsilon_{0}$ is the vacuum dielectric constant, $\hat{{n}}$ is the interface normal direction, and $\sigma_{\hat{{n}}}$ is the interfacial density of  bond dipoles projected onto the $\hat{{n}}$ direction, 
\begin{equation}
	\sigma_{\hat{{n}}} = \frac{\vec{{p}} \cdot \hat{{n}}}{S_{\hat{{n}}}},
\end{equation} 
where $S_{\hat{{n}}}$ is the area per bond in the interface plane. For any interface direction, the $V_{D_{R}}^{\hat{{n}}}$ can be obtained from an atomistic DFT supercell calculation, by taking the difference between the band offset of electronically relaxed interface and that of the Wigner-Seitz interface. Then the magnitude of $\vec{{p}}$ is given by $|\vec{{p}}|=\varepsilon_{0} V_{D_{R}}^{\hat{{n}}} S_{\hat{{n}}} / |\cos(\theta)|$ where $\theta$ is the angle between $\vec{{p}}$ and $\hat{{n}}$.

Based on this, we calculate the magnitude of $\vec{{p}}$ implied by the band offset of Si-Ge along the different orientations in Figs. 3(a-d). We note that even the interfacial bonds can have more than one direction (e.g., two for [100] interface), they are symmetrically equivalent and hold the same $|\cos(\theta)|$, thus the above expression of $|\vec{{p}}|$ still works. For the [100] and [110] interfaces of Si-Ge, $|\vec{{p}}|$ is calculated to be 26.11$\times 10^{-3}$ e$\cdot$\AA\ and 26.08$\times 10^{-3}$ e$\cdot$\AA\, respectively. For the [111] direction, the interface can have one or three interfacial bonds per unit cell area, and $|\vec{{p}}|$ is calculated to be 26.13$\times 10^{-3}$ e$\cdot$\AA\ and 26.01$\times 10^{-3}$ e$\cdot$\AA, respectively. This independence of $|\vec{{p}}|$ on interface orientation (to within 0.4\%)  is consistent with our basic observation of highly localized bond dipoles distributed along the interfacial bonds. 

Given this simple model of electron relaxation at the interface, we revisit the series interface VBOs investigated in Fig. 2. Figure 2(b) shows the results wherein the potential shift due to the presence of bond dipoles (BD) is added to WSA results. Here the magnitude $|\vec{{p}}|$, which is assumed as a materials-specific constant for each interface system, is calculated using the [110] direction for all 3D systems, zigzag direction for 2D hexagonal, and [11] direction for 2D square lattices. Such determined $|\vec{{p}}|$ is used together with Eqs. (5-6) to calculate the VBO including the electron relaxation induced interfacial dipole, dubbed as the WSA+BD model. Here we note that while the BD is explicitly calculated along the specified direction, it is transferrible and the reported WSA+BD values are for orientations excluding the orientation for which it is calculated (which would definitionally yield exact results). In Fig. 2(b), the inclusion of the bond dipole contributions almost perfectly reproduces the atomistic DFT results with root mean square deviation of 27 meV, which is already close to the error introduced when using the average potential alignment to determine the VBO from a supercell in DFT calculation (readout error). This agreement indicates that the band offset for different directions can be understood in terms of the magnitude of the bond dipole and the geometry of the interface.

While Eq. (6) is given for a case where all bond dipoles are identical, for more general cases or even anisotropic systems, one has to explicitly consider the contributions of different bond directions, i.e. $\hat{{d}} \in  \{ \hat{{d}}_{i} \}$, where the index $i$ is generally over different types of bonds as well as the bond orientations. We then have to sum up the contributions of $\{ \hat{{d}}_{i} \}$ to calculate the final results of $\sigma_{\hat{{n}}}$. This is illustrated by a 2D triangular lattice in Figs. 3(e-g), where the three different bond directions are indicated by green $\hat{{d}}_{1}$, blue $\hat{{d}}_{2}$ and red $\hat{{d}}_{3}$ arrows. To facilitate the counting of bond dipoles, we note that for each $\hat{{d}}_{i}$, all the bonds of that orientation line up end-to-end, forming a family of ``bond lines'' (the solid lines in the same color of $\hat{{d}}_{i}$). The formation of a 2D interface along the thin black line in Figs. 3(e-g) can be seen as the truncation of bond lines which will then be reconnected to a dissimilar material. Here the key point is that the interface crosses every bond line \emph{once and only once}. Thus, \textit{assuming that cutting each bond line corresponds to the cutting of a chemical bond}, each bond line will contribute a corresponding bond dipole $\vec{{p}}_i$. The spacing between $\hat{{d}}_{i}$ bond lines along the interface is denoted $S_{\hat{{n}}}(\hat{{d}}_{i})$ and is highlighted by the thick black line in Figs. 3(e-g). The contribution of $\hat{{d}}_{i}$ bonds to the projected interfacial dipole density Eq. (6) is $\vec{{p}}_i \cdot \hat{{n}}/ S_{\hat{{n}}}(\hat{{d}}_{i})$. By summing up the contributions of all the different $\{ \hat{{d}}_{i} \}$, the total interfacial dipole density can then generally be expressed as  
\begin{equation}
	\sigma_{\hat{{n}}} =\sum_{i} \frac{\vec{{p}}_{i} \cdot \hat{{n}}}{S_{\hat{{n}}}(\hat{{d}}_{i})} =\sum_{i} |\vec{{p}}_{i}|\frac{\hat{{d}}_{i} \cdot \hat{{n}}}{S_{\hat{{n}}}(\hat{{d}}_{i})}, 
\end{equation} 
where $|\vec{{p}}_{i}|$ is the magnitude of bond dipole, which for a single type of bond would just be a constant, $p$. The direction of the resultant bond dipole is along the bond lines, where the positive direction has been chosen such that $\vec{{p}}_{i} = |\vec{{p}}_{i}|\hat{{d}}_{i}$.

While the magnitude of bond dipole $|\vec{{p}}_{i}|$ of Eq. (7) is independent of interface orientation, it is not obvious how orientation independence arises in $\sigma_{\hat{{n}}}$ as both $\hat{{d}}_{i} \cdot \hat{{n}}$ and the bond line spacing $S_{\hat{{n}}}(\hat{{d}}_{i})$ depend explicitly on the interface direction $\hat{{n}}$. In Fig. 3(e), the bond lines associated with the $\hat{{d}}_{1}$ direction are shown with green lines and the interface in the $\hat{{n}}$ direction is indicated by the thin black line. The length of interface per bond line, $S_{\hat{{n}}}(\hat{{d}}_{1})$, is indicated by the thick black line. The perpendicular distance between bond lines is given when $\hat{{n}}=\hat{{d}}_{1}$ which we write as $S_{\hat{{d}}_{1}}$ (instead of  $S_{\hat{{n}}=\hat{{d}}_{1}}(\hat{{d}}_{1})$ for brevity).  Here, it can be seen that $S_{\hat{{n}}}(\hat {{d}}_{1})= S_{\hat{{d}}_{1}}/\cos(\theta)$, where $\theta$ is the angle between $\hat{{n}}$ and $\hat{{d}}_{1}$, or more generally $S_{\hat{{n}}}(\hat {{d}}_{i})= S_{\hat{{d}}_{i}}/\cos(\theta_i)$. Note, this relationship can be straightforwardly generalized to 3D, as shown in Figs. 3(h-i) where the plane perpendicular to the bond lines, $S_{\hat{{d}}_{1}}$, is shown in blue and the interface plane ${S_{\hat{{n}}}(\hat{{d}}_1)}$ is shown in black. As $\hat{{n}}$ and $\hat{{d}}_i$ are the normal vectors for  $S_{\hat{{n}}}(\hat{{d}}_i)$  and  $S_{\hat{{d}}_i}$, respectively,  the general expression for ${S_{\hat{{n}}}(\hat{{d}}_i)}$ can be written as, 
\begin{equation}
{S_{\hat{{n}}}(\hat{{d}}_i)}= \frac{S_{\hat{{d}}_i}}{|\cos(\theta_i)|} =\frac{S_{\hat{{d}}_i}}{|\hat{{n}} \cdot \hat{{d}}_i|},
\end{equation}
where $\theta_i$ the angle between $\hat{{n}}$ and $\hat{{d}}_i$. Thus,  $\sigma_{\hat{{n}}}$ has the formula

\begin{equation}
	\sigma_{\hat{{n}}} = \sum_{i} \frac{|\vec{{p}}_i| }{S_{\hat{{d}}_i}} |\hat{{d}}_i \cdot \hat{{n}}|^2.
\end{equation}

For isotropic structures of elemental and binary systems in Table I, both $|\vec{{p}}_i|$ and $S_{\hat{{d}}_i}$ are independent of $\hat{{d}}_i$ so that $|\vec{{p}}_i| = p$ and $S_{\hat{{d}}_i} = S$. As discussed below, for such systems it can be explicitly shown that the summation in Eq. (9) is independent of $\hat{{n}}$, simplifying to 
\begin{equation}
	\sigma_{\hat{{n}}} = \frac{p}{S} \sum_{i}  |\hat{{d}}_i \cdot \hat{{n}}|^2 = \frac{p}{S} C=\sigma,
\end{equation} 
where $C= \sum_{i}  |\hat{{d}}_i \cdot \hat{{n}}|^2$ sums over the projection of different bond lines on the normal direction $\hat{{n}}$. In general, for a system with $N$-fold symmetry,
\begin{equation}
	C = \frac{N}{D},
\end{equation}
where $D$ is the system dimensionality ($D=2$ or $3$).

In 2D, for the lattice to be isotropic, it must have an $N$-fold rotational symmetry with $N \geq 3$. The whole set of bond directions $\{ \hat{{d}}_{1}, \hat{{d}}_{2}, ..., \hat{{d}}_{N} \}$ will equally partition a circle into $N$ parts, thus each bond direction $\hat{{d}}_{i}$ has a relative angle of $\frac{2\pi}{N}$ from the previous one $\hat{{d}}_{i-1}$. If we take the angle between $\hat{{d}}_{1}$ and $\hat{{n}}$ to be $\theta$, 
\begin{equation}
	\begin{aligned}
	C &= \sum_{i=1}^{N} \left|\cos \left(\theta + (i-1)\frac{2\pi}{N}\right)\right|^2 \\
	  &= \sum_{i=1}^{N} \frac{1}{4}\left[ e^{i\left(\theta + (i-1)\frac{2\pi}{N}\right)}+e^{-i\left(\theta + (i-1)\frac{2\pi}{N}\right)} \right]^2.
	\end{aligned}
\end{equation}
By taking the summation over $i$, one can get
\begin{equation}
	C = \frac{N}{2} + e^{i2\theta}\frac{1-1}{1-e^{i\frac{4\pi}{N}}} + e^{-i2\theta}\frac{1-1}{1-e^{-i\frac{4\pi}{N}}},
\end{equation}
which leads to $C = \frac{N}{2}$ for $N \geq 3$. Note that $N=2$ is associated to 1D linear symmetry and not considered here.

In 3D, the general $N$-fold equal partition of a sphere is still an active area of research\cite{gorshi2005}. However, for conventional semiconductors $N$ is limited to only several possibilities, e.g. 3 (simple cubic), 4 (bcc, diamond), or 6 (fcc).  For a simple cubic lattice, for example, $\hat{{d}}_{i}$'s are explicitly,
\begin{equation}
	\begin{aligned}
	\hat{{d}}_{1} &= \left(1, 0, 0\right) \\
	\hat{{d}}_{2} &= \left(0, 1, 0\right) \\
    \hat{{d}}_{3} &= \left(0, 0, 1\right).
    \end{aligned}
\end{equation}
For an arbitrary $\hat{{n}} = \left(\alpha, \beta, \gamma\right)$, we can explicitly see that 
\begin{equation}
C=\sum_{i} |\hat{{d}}_i \cdot \hat{{n}}|^2 =\alpha^2+\beta^2+\gamma^2=\frac{3}{3},
\end{equation}
noting that $\hat{{n}}$ is a normalized unit vector such that $\alpha^2+\beta^2+\gamma^2=1$.
In Note S3 in the SM, $C$ is explicitly calculated for each isotropic lattice, in which we find that for 3D systems $C = \frac{N}{3}$. Taking the results of 2D and 3D together, $C = \frac{N}{D}$ is an orientation independent constant determined by the symmetry and dimension of the system.

While Eq. (10) holds for most of the isotropic lattices, some further consideration has to be taken for 2D hexagonal and 3D diamond/zincblende structures, in which our assumption that cutting a bond line is equivalent to cutting a chemical bond breaks down. Despite the interface crossing each bond line exactly once, only cutting an actual chemical bond will yield a bond dipole which contributes to the band offset. This raises no problems for most of the isotropic lattices since, as exemplified in Figs. 3(e-g), the bond lines are \emph{fully-filled} by chemical bonds so that the interface will always cut a chemical bond on each bond line. However, in 2D hexagonal and 3D diamond/zincblende lattices, for a given bond line, whether or not a chemical bond is cut depends on the absolute position of the cutting plane, since the bond lines are only \emph{partially-filled} by chemical bonds. Thus, only a finite fraction of bond lines contribute bond dipoles. In such case, as detailed in Note S4 in the SM, the cutting of each bond line $\hat{{d}}_{i}$ effectively contributes a corresponding dipole $\lambda_i \vec{{p}}_i$, where $\lambda_i$ is the``filling factor'' and indicates what fraction of the bond line is occupied by actual chemical bonds. Thus, Eq. (9) can be generalized as 
\begin{equation}
	\sigma_{\hat{{n}}} = \sum_{i} \frac{\lambda_i |\vec{{p}}_i| }{S_{\hat{{d}}_i}} |\hat{{d}}_i \cdot \hat{{n}}|^2.
\end{equation}
For isotropic lattices, $\lambda_i$ is also independent of $\hat{{d}}_{i}$ such that
\begin{equation}
	\sigma_{\hat{{n}}} = \frac{\lambda p}{S} C=\sigma,
\end{equation} 
where 2D hexagonal and 3D diamond/zincblende lattices have $\lambda = \frac{1}{3}$ and $\frac{1}{4}$, respectively, while in other cases $\lambda$ simply equals to unity.  

These results indicate that the interfacial dipole which develops at the interface can be well described by point dipoles which exist at the bonds between atoms at the interface. For an interface between isotropic monoatomic materials, this leads to an interfacial dipole which is strictly independent of the interface orientation. While in our treatment we have envisioned an atomically flat interface, it is noteworthy that this result holds even when considering a rough interface. As shown in Note S5 in the SM, altering the shape of the interface between two monoatomic materials yields no change in the band offset for any particular interface orientation. In general, however, for binary AX semiconductors, different types of interfacial bonds exist. As such, surfaces can be A or X terminated, leading to polar interfaces, which is the subject of our next section.

\begin{table}
	\centering
	\caption{Material constants defined in Eqs. (10) and (17) for different lattice structures. The lattice constant of traditional unit cell is set to $a$.}
	\begin{tabular}{lccccccccccccccccccccccccccc}
		&Dimension      &Lattice      &$\lambda$           &$S$                       &$N$-fold      &$C$\\
		\hline			
		&3D    &simple cubic          &$1$                 &$a^2$                     &3             &$\frac{3}{3}$ \\
		& \\
		&      &body-centered cubic   &$1$                 &$\frac{\sqrt{3}}{6}a^2$   &4             &$\frac{4}{3}$ \\
		& \\
		&      &face-centered cubic   &$1$                 &$\frac{\sqrt{2}}{4}a^2$   &6             &$\frac{6}{3}$ \\
		& \\
		&      &rock-salt             &$1$                 &$\frac{1}{4}a^2$          &3             &$\frac{3}{3}$ \\
		& \\
		&      &cesium chloride      &$1$                  &$\frac{\sqrt{3}}{6}a^2$   &4             &$\frac{4}{3}$ \\
		& \\
		&      &diamond/zincblende    &$\frac{1}{4}$       &$\frac{\sqrt{3}}{12}a^2$  &4             &$\frac{4}{3}$ \\
		& \\
        \hline
		&2D    &square                &$1$                 &$a$                       &2             &$\frac{2}{2}$ \\
		& \\
   		&      &triangular            &$1$                 &$\frac{\sqrt{3}}{2}a$     &3             &$\frac{3}{2}$ \\
        & \\
        &      &hexagonal             &$\frac{1}{3}$       &$\frac{1}{2}a$            &3             &$\frac{3}{2}$ \\
        & \\
        \hline
	\end{tabular}
\end{table}

\subsection{Non-polar and polar interfaces}

\begin{figure}[tbp]
	\includegraphics[width=\columnwidth]{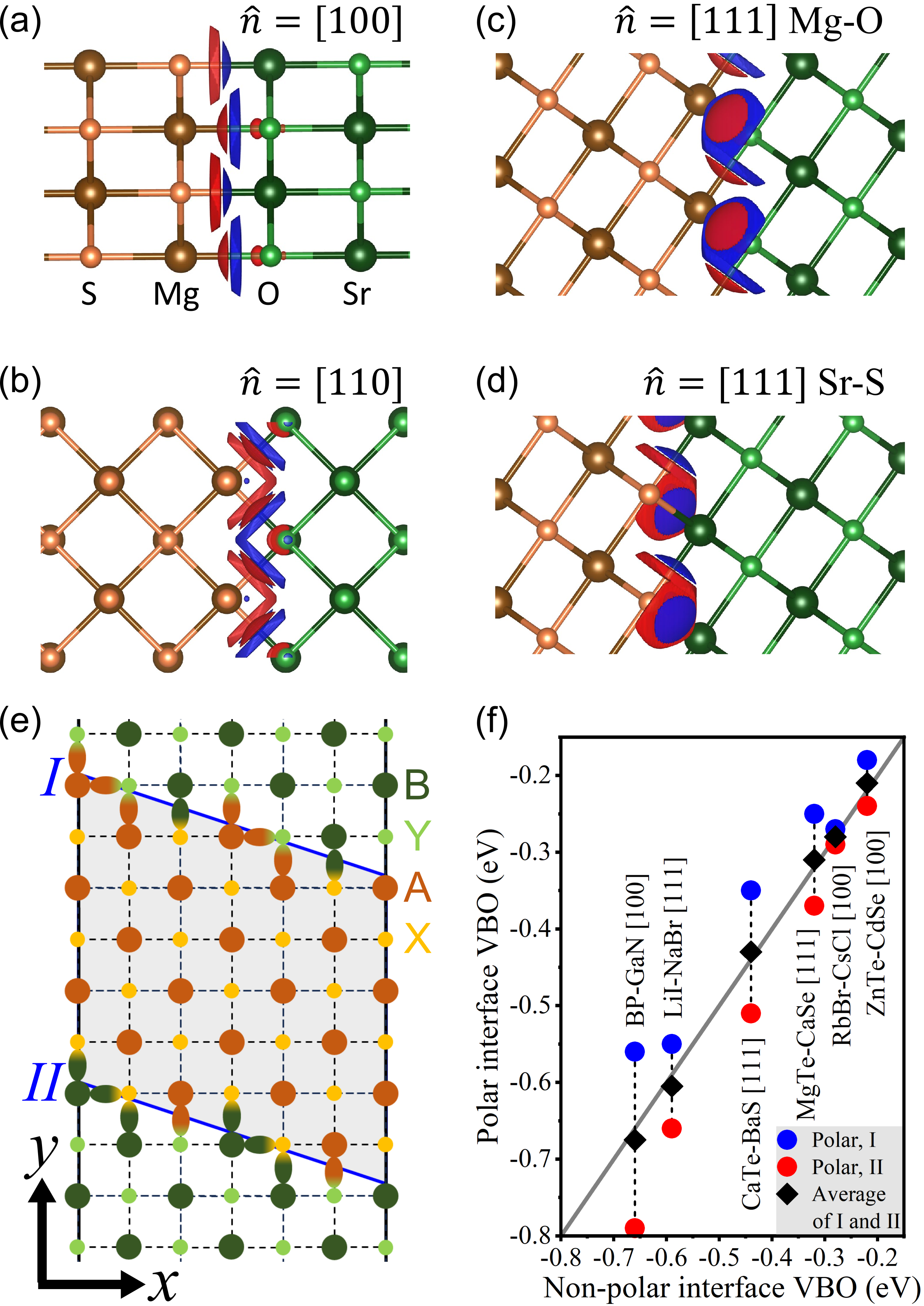}
	\caption{\label{fig:fig4} \textbf{Interfacial bond dipole in the binary system.} (a-d) The charge distribution of bond dipoles at MgS(left)-SrO(right) interface along different orientations: (a) [100], (b) [110], (c) [111] with Mg-O bond, and (d) [111] with Sr-S bond. Red (blue) cloud denotes the positive (negative) dipole charges. (e) Schematic of the interfacial bonds in a polar direction in 2D checkerboard lattice. The solid blue lines denote two complementary polar interfaces with opposite $\eta^{A}_{i}$ and $\eta^{B}_{i}$. Two types of interfacial bonds are shown by ellipses in bi-colors associated to the atoms forming the bond. (f) The atomistic DFT results of VBO for 6 different polar interfaces (details in Table S2 in the SM). Horizontal (vertical) axis is the VBO of non-polar (polar) interface. Grey line indicates the condition that the VBO of polar and non-polar interface are equal. Blue (red) points denote the ``Polar, I'' (``Polar, II'') interface and the black diamonds shows the average value of I and II.}
\end{figure}

The surface of a binary can be either non-polar or polar, of which the essential difference is that non-polar surface contains an equal number of cations and anions while for a polar surface their numbers are different. When reconnecting two dissimilar binaries, two types of interfacial bonds can form depending on how the surfaces are terminated. Considering two binaries AX and BY with A, B (X, Y) being the cations (anions), both A-Y and B-X bonds can form at the interface. The key point is that a non-polar interface contains two types of interfacial bonds with equal populations, while the relative proportions of the two bond types are different for polar interfaces. This is exemplified using MgS-SrO interface shown in Figs. 4(a-d), where (a-b) are non-polar interfaces with equal numbers of Mg-O and Sr-S bonds while (c) and (d) are polar interfaces containing only Mg-O or Sr-S bonds, respectively. While these are limiting cases, in general we can define $\eta^{A}_{i}$ ($\eta^{B}_i$) which are the relative concentrations of bond-type A (B) in $\hat{{d}}_i$ direction at the interface wherein $\eta^{A}_{i}+\eta^{B}_{i}=1$ for all bond directions. As the basic geometric properties of the interface are unchanged for the case of binaries, we can then generalize Eq. (16) as
\begin{equation}
	\sigma_{\hat{{n}}} = \sum_{i} \frac{\lambda_i}{S_{\hat{{d}}_i}} \left(\eta^{A}_{i}|\vec{{p}}^{A}_{i}| + \eta^{B}_{i}|\vec{{p}}^{B}_{i}|\right) |\hat{{d}}_i \cdot \hat{{n}}|^2,
\end{equation}
where $\vec{{p}}^{A}_{i}$ and $\vec{{p}}^{B}_{i}$ are the bond dipoles associated with the two types of interfacial bonds (i.e., associated with A-Y and B-X bonds, respectively) in $\hat{{d}}_i$ direction.  Despite these two types of bonds, we can see that for any non-polar interface $\eta^{A}_{i}=\eta^{B}_{i}=\eta=\frac{1}{2}$ and by taking $p =\left(|\vec{{p}}^{A}_{i}| + |\vec{{p}}^{B}_{i}|\right)/2$, the orientation independence given by Eq. (17) is strictly restored.

For polar interfaces, however, we note that the concentrations $\eta^{A}_{i}$ and $\eta^{B}_{i}$ depend not only on orientation, but also on the absolute position of the interfacial plane wherin a shift of the interface will be associated with different terminating atoms in each material and hence lead to different types of polar interfaces.  Taking MgS-SrO [111] as an example, two different interfaces can be constructed as illustrated in Figs. 4(c-d), where the interfacial bonds can be either Mg-O or Sr-S depending on the termination. These two polar interfaces can be described by ($\eta^{A}_{i}=1$, $\eta^{B}_{i}=0$) and ($\eta^{A}_{i}=0$, $\eta^{B}_{i}=1$), with interfacial dipole density of $\sigma_{\hat{{n}}}^I = \sum_{i} \frac{\lambda_i}{S_{\hat{{d}}_i}} |\vec{{p}}^{A}_{i}| |\hat{{d}}_i \cdot \hat{{n}}|^2 $ and $\sigma_{\hat{{n}}}^{II} = \sum_{i} \frac{\lambda_i}{S_{\hat{{d}}_i}} |\vec{{p}}^{B}_{i}| |\hat{{d}}_i \cdot \hat{{n}}|^2 $, respectively. We deem these two interfaces as complimentary as $\eta_i$ is replaced by the complementary $1 - \eta_i$ and note that provided the two materials are isostructural, such a complimentary interface exists (as can be confirmed by swapping the two materials). A more concrete example of such a pair of interfaces is shown in Fig. 4(e) with the 2D checkerboard lattice where two complementary interfaces are shown by solid blue lines. Here the two bond line directions are $\hat{{d}}_{i=x}$ and $\hat{{d}}_{i=y}$ along x- and y-axis. The two different polar interfaces have complementary relative concentrations ($\eta^{A}_{i=x}=1$, $\eta^{A}_{i=y}=\frac{2}{3}$, $\eta^{B}_{i=x}=0$, $\eta^{B}_{i=y}=\frac{1}{3}$) for $I$ and ($\eta^{A}_{i=x}=0$, $\eta^{A}_{i=y}=\frac{1}{3}$, $\eta^{B}_{i=x}=1$, $\eta^{B}_{i=y}=\frac{2}{3}$) for $II$. Indeed, we see that the summation of the relative concentrations of two complementary polar interfaces equals to twice of the value of a non-polar interface, 
\begin{equation}
\sigma_{\hat{{n}}}^{NP}= (\sigma_{\hat{{n}}}^I + \sigma_{\hat{{n}}}^{II})/2.
\end{equation}
As such, the orientation independence of band offset in polar directions can be understood from their average effect. This relation is also demonstrated from direct calculation in Fig. 4(f), where the VBO associated to $\sigma_{\hat{{n}}}^{I(II)}$ are shown in blue(red) and their average value (black) is found to be quite close to the non-polar interfaces indicated by the grey line. Note that in Fig. 4 we only consider the polar interfaces formed by isovalence materials as heterovalent ions across a polar interface are expected to donate/accept charge from bulk \textemdash leading to extended states which will not be well represented by point bond dipoles.

We can see that the band offset of a single polar interface is different from that of non-polar directions and depends on the atomic termination of interface plane. As an example\cite{funato1999}, it is experimentally seen that the VBO of GaAs-ZnSe [100] interface can be tuned from 0.6 to 1.1 eV by controlling the atomic configuration of interface. However, more experimental interests are usually given to a special case that two constituent materials of the interface have a common anion/cation, e.g., AlAs-GaAs\cite{hirakawa1990} and CdTe-HgTe\cite{becker1991}, where the band offset is found to be consistent between polar and non-polar interface. In such cases, two complementary polar interfaces becomes equivalent under a lattice translation along the interface normal direction. This symmetry forces two types of bond dipoles $\vec{{p}}^{A}_{i}$ and $\vec{{p}}^{B}_{i}$, which although are associated to different interfacial bonds, to have the same magnitude $|\vec{{p}}^{A}_{i}| = |\vec{{p}}^{B}_{i}| = p$. Thus, $\sigma_{\hat{{n}}}$ is single-valued for a common-anion/cation interface even in the polar directions and the orientation independence of band offset is rigorously restored for \emph{any} orientation. 

\subsection{Band alignment in anisotropic systems}

\begin{figure}[tbp]
	\includegraphics[width=\columnwidth]{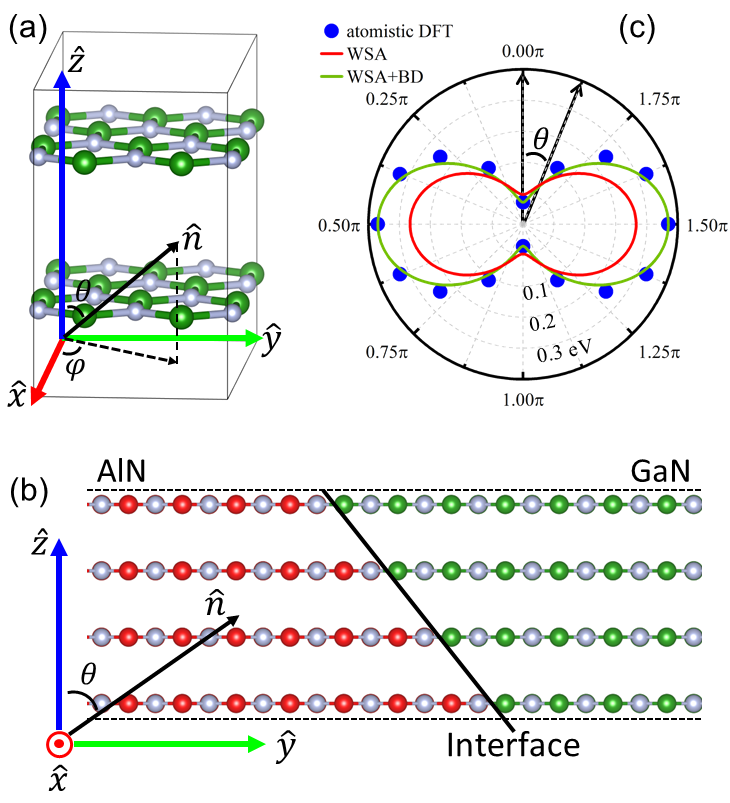}
	\caption{\label{fig:fig5} \textbf{Band offset of AlN-GaN heterostructure.} (a) Lattice Schematic of the vdW bulk of AA stacking hexagonal layers. The normal vector $\hat{{n}}$ of any surface can be presented by two angles $\theta$ and $\varphi$. (b) Supercell structure of the interface between two vdW bulk materials AlN and GaN. Without loss of generality, we set $\varphi=90^{\circ}$ and the normal vector $\hat{{n}}$ can rotate in $y-z$ plane by changing $\theta$. The interface plane is shown as the solid black line. (c) Calculated VBO of AlN-GaN interface in (b) as a function of $\theta$ with a fixed $\varphi=90^{\circ}$. Blue dots denote the atomistic DFT results. Red line shows the result of Wiger-Seitz atomic interface. Green line shows the result of Eq. (26), which includes both bulk quadrupole and bond dipole contributions.}
\end{figure}

In this section we will apply the bond dipole model to the interface between two anisotropic systems to investigate how well it describes the orientation dependence of the band alignment. As a protypical example we consider  the interface between van der Waals (vdW) bulk materials, AlN and GaN, which are both layered hexagonal lattices with AA stacking. We note that both of these materials have a strong in-plane chemical bonding and a weak out-of-plane vdW binding, and thus have significant anisotropy. An interface orientation between these two materials can be descibed by its normal $\hat{{n}}$, 

\begin{equation}
	\begin{aligned}
		\hat{{n}} = \left(\alpha, \beta, \gamma\right) = \left(\sin\theta\cos\varphi, \sin\theta\sin\varphi, \cos\theta\right), 
	\end{aligned}
\end{equation}
where $\theta$ and $\varphi$ are the zenith and azimuthal angles as shown in Fig. 5(a).
Writing the quadrupole potential shift ${V}_{vac}^{\hat{{n}}} - \overline{V}_{bulk}$ of Eq. (2) in terms of $\theta$ and $\varphi$, we obtain
\begin{equation}
	\begin{aligned}
		{V}_{vac}^{\hat{{n}}} - \overline{V}_{bulk} = \frac{2\pi}{\Omega} \sum_{i} &\int d^{3}{r}' [\sin^2\theta\cos^2\varphi {x'}^2 + \\ &\sin^2\theta\sin^2\varphi {y'}^2 + \cos^2\theta {z'}^2 ] \rho_{i}\left({\vec{r}}'\right).
	\end{aligned}
\end{equation}
Note that there are no cross terms in Eq. (21) due to the 2D isotropy of hexagonal lattice (eliminating $xy$) and the mirror symmetry of AA stacking (eliminating $xz$ and $yz$). As $\theta$ and $\varphi$ are constants in the intergration, evaluating Eq. (21) yields
\begin{equation}
	\begin{aligned}
		{V}_{vac}^{\hat{{n}}} - \overline{V}_{bulk} = \frac{2\pi}{\Omega} [ &\sin^2\theta\cos^2\varphi Q^{\hat{{x}}\hat{{x}}} + \sin^2\theta\sin^2\varphi Q^{\hat{{y}}\hat{{y}}} \\ &+ \cos^2\theta Q^{\hat{{z}}\hat{{z}}} ],
	\end{aligned}
\end{equation}
where $Q^{\hat{{x}}\hat{{x}}}$, $Q^{\hat{{y}}\hat{{y}}}$ and $Q^{\hat{{z}}\hat{{z}}}$ are the sum of quadrupoles of the Wigner-Seitz atoms in unit cell along the $\hat{{x}}$, $\hat{{y}}$ and $\hat{{z}}$ directions. Using Eq. (1), Eq. (22) can also be rewritten as,
\begin{equation}
	\begin{aligned}
		{V}_{vac}^{\hat{{n}}} - \overline{V}_{bulk} = &\sin^2\theta\cos^2\varphi ({V}_{vac}^{\hat{{x}}} - \overline{V}_{bulk}) +\\ &\sin^2\theta\sin^2\varphi ({V}_{vac}^{\hat{{y}}} - \overline{V}_{bulk}) +\\ &\cos^2\theta({V}_{vac}^{\hat{{z}}} - \overline{V}_{bulk}).
	\end{aligned}
\end{equation}
The band offset of a Wigner-Seitz interface is obtained by taking the difference of Eq. (23) between two dissimilar materials across the interface. Note here that the bulk quadrupole contribution to the band offset is orientation dependent because the quadrupole is not isotropic.

To simplify the results, we focus on the band offset of non-polar interfaces (or alternately the average of two complementary polar interfaces). Similar to Eqs. (18-19), for non-polar interfaces there are an equal number of Al-N and Ga-N bond dipoles and hence they can be grouped into an average dipole. For the stacked vdW material, there exist three different bond directions $\hat{{d}}_1$, $\hat{{d}}_2$ and $\hat{{d}}_3$, associated with the strong in-plane covalent bonds of the hexagonal lattice. Additionally, there is weaker out-of-plane vdW binding between layers which may contribute to the bond dipole at the interface, with an associated bond line direction $\hat{{d}}_{4}$. Thus, for a vdW hexagonal lattice we have $\hat{{d}} \in  \{ \hat{{d}}_{1},\hat{{d}}_{2},\hat{{d}}_{3},\hat{{d}}_{4} \}$.     

From Eq. (16), we then have
\begin{equation}
	\begin{aligned}
	\sigma_{\hat{{n}}} = &\left( \sum_{i} \frac{\lambda_{i} |\vec{{p}}_{i}| }{S_{\hat{{d}}_i}} |\hat{{d}}_i \cdot \hat{{x}}|^2 \right) \sin^2\theta\cos^2\varphi +\\ &\left( \sum_{i} \frac{\lambda_{i} |\vec{{p}}_{i}| }{S_{\hat{{d}}_i}} |\hat{{d}}_i \cdot \hat{{y}}|^2 \right) \sin^2\theta\sin^2\varphi +\\ &\left( \sum_{i} \frac{\lambda_{i} |\vec{{p}}_{i}| }{S_{\hat{{d}}_i}} |\hat{{d}}_i \cdot \hat{{z}}|^2 \right) \cos^2\theta,
	\end{aligned}
\end{equation}
where the terms in parenthesis are the interfacial dipole density for an interface with the normal direction in $\hat{{x}}$, $\hat{{y}}$ and $\hat{{z}}$, respectively. Thus, $\sigma_{\hat{{n}}}$ can be expressed as
\begin{equation}
	\sigma_{\hat{{n}}} = \sigma_{\hat{{x}}} \sin^2\theta\cos^2\varphi + \sigma_{\hat{{y}}} \sin^2\theta\sin^2\varphi + \sigma_{\hat{{z}}} \cos^2\theta,
\end{equation}
and combining Eqs. (23) and (25), the VBO in any direction $\hat{{n}}$ is given by
\begin{equation}
	\begin{aligned}
	VBO(\hat{{n}}) = &VBO(\hat{{x}}) \sin^2\theta\cos^2\varphi +\\ &VBO(\hat{{y}}) \sin^2\theta\sin^2\varphi +\\ &VBO(\hat{{z}}) \cos^2\theta.
	\end{aligned}
\end{equation}
Eq. (26) takes into account both bulk quadrupole and bond dipole contributions to the band offset with the only assumption that the bond dipoles are unaffected by the interfacial orientation.

In Fig. 5(b), without loss of generality, we consider the AlN-GaN interface with $\varphi=90^{\circ}$ and normal direction $\hat{{n}}$ changes as a function of $\theta$. Since the hexagonal lattice is isotropic in 2D, changing $\varphi$ does not affect the result of band offset. In Fig. 5(c), we compare the VBO calculated from the full atomistic DFT for AlN-GaN interface (shown as blue dots) to those predicted from the Wigner-Seitz atom interface (red line) and the Wigner-Seitz atom interface where we have explicitly included the contribution of the bond dipole presented in Eq. (26) (green line). Here we see that while the qualitative features are somewhat captured by WSA, WSA+BD shows dramatic improvement and quantitative agreement with the DFT results with a root mean square deviation of only 28 meV, nearly identical to the deviations observed in the isotropic case.

\subsection{Extension to alloy systems}

In extending to alloy systems, we consider a pure isotropic material B in which isovalent impurities of type A have been introduced to yield $A_xB_{1-x}$. While any particular arrangement of impurities may break the isotropy of bulk, it is nevertheless preserved by the overall ensemble average. A correct theory of the band alignment of alloy must guarantee the \emph{statistical} orientation independence of the average potential, for which the ``Wigner-Seitz atom + bond dipole'' model will be a straightforward choice.

From Eq. (1) the quadrupole, $\hat{{n}}^{T} \overleftrightarrow{{Q}} \hat{{n}}$, determines the bulk average potential in $\hat{{n}}$ direction relative to the vacuum level. For isotropic systems, $\hat{{n}}^{T} \overleftrightarrow{{Q}} \hat{{n}}$ is generally a constant under rotation, 
\begin{equation}
	\begin{aligned}
		&\hat{{n}}^{T} \overleftrightarrow{{Q}} \hat{{n}} =  \sum_{i} \int_\Omega d^{3}{r} \left(\hat{{n}} \cdot {\vec{r}}\right)^2 \rho_{i}\left({\vec{r}}\right) \\
		&= \sum_{i} \int_\Omega d^{3}{r} \left[ \left(\hat{{n}} \cdot {\vec{r}}\right)^2 - \frac{r^2}{3} \right] \rho_{i}\left({\vec{r}}\right) + \sum_{i} \int_\Omega d^{3}{r} \frac{r^2}{3} \rho_{i}\left({\vec{r}}\right).
	\end{aligned}
\end{equation}
The first term in Eq. (27) is the traceless quadrupole which has explicit direction dependence and thus must vanish for the case of isotropic systems. Therefore, the average potential of isotropic systems is governed entirely by the average trace of the quadrupole $\int_\Omega d^{3}{r} \frac{1}{3}r^2 \rho_{i}\left({\vec{r}}\right)$, which has been deemed the spherapole in Ref.\cite{tung2014}, despite the lack of spherical symmetry in any crystal.

As a result, the alloy system becomes isotropic and the band offset is governed entirely by the ensemble average of the spherapole term. Considering a large unit cell of an alloy consisting of two types of atoms A and B, this spherapole term becomes 
\begin{equation}
\sum_{i} \int_\Omega d^{3}{r} \frac{r^2}{3} \rho_{i}\left({\vec{r}}\right)= n_A  Q_A + n_B  Q_B,
\end{equation}
with $n_A$ and $n_B$ the number of A and B atoms. Here we have assumed that there is no charge transfer between the Wigner-Seitz atoms of A and B, which could modify $Q_A$ or $Q_B$ depending on chemical environment. Considering charge transfer between these atoms discussed in the Section C, a more thorough accounting of the total quadrupole would include, for example the shift of $Q_B$ relative to $Q_A$ due to local charge transfer. Assuming the charge transfer depends only on local chemical bonding but not long-range structure, Eq. (27) might be generalizable to other statistically isotropic systems such as amorphous materials or even liquids.

\subsection{Contribution of ionic relaxation}
\begin{figure}[tbp]
	\includegraphics[width=\columnwidth]{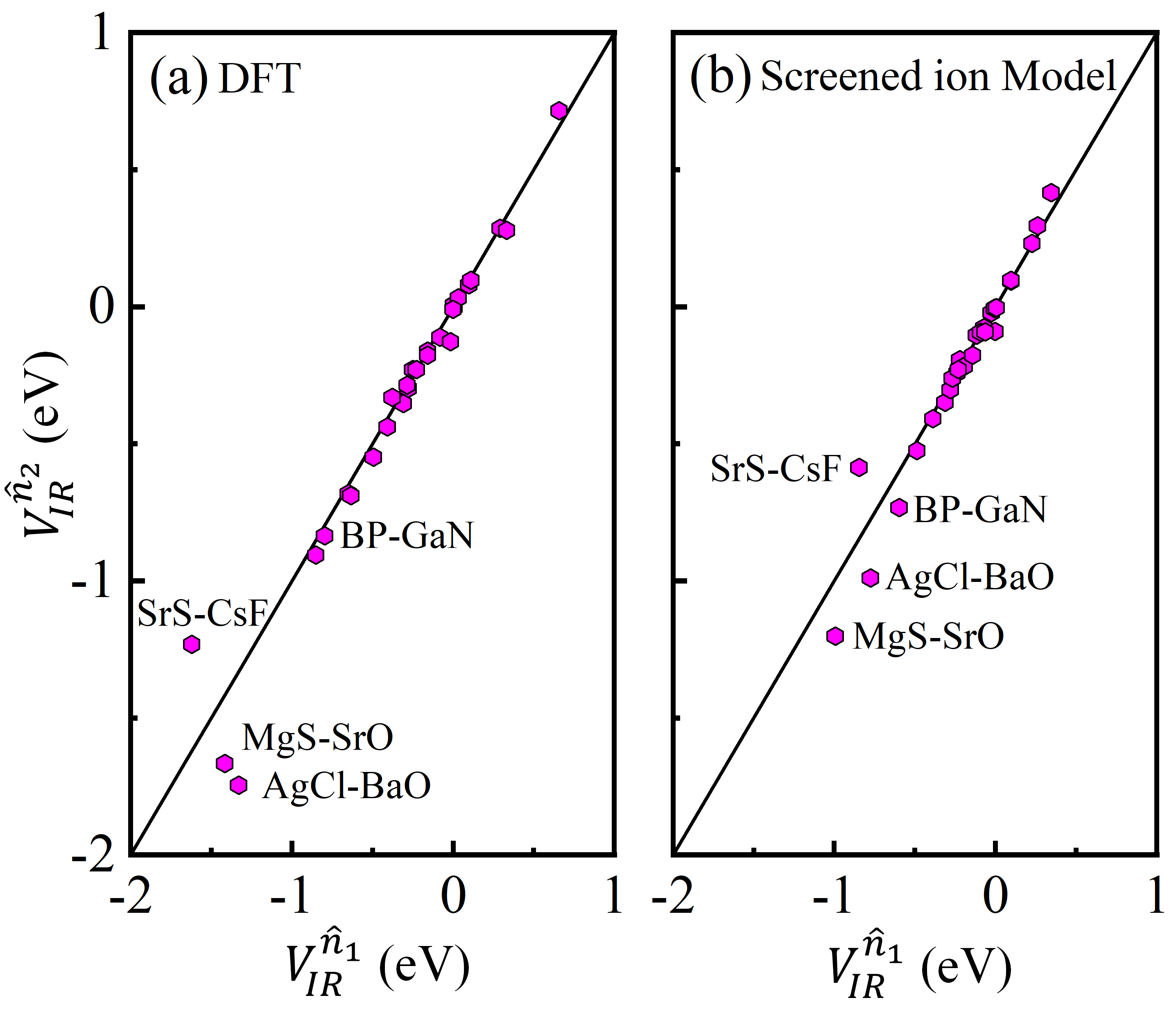}
	\caption{\label{fig:fig6} \textbf{Ionic contribution to VBO.} The shift of VBO due to ionic relaxation in two different non-polar directions $\hat{{n}}_1$ and $\hat{{n}}_2$ calculated using (a) DFT and (b) the screened ion model. In both panel the black line indicates the case that horizontal and vertical axes are equal.}
\end{figure}

The results presented thus far and our bond dipole theory are associated with electron relaxation at the interface and this electron contribution to the band offset. For completeness, we also report here the contribution of ionic relaxation to the band offset. We calculate the shift of VBO due to ionic relaxation, $V_{IR}^{\hat{{n}}}$, by first calculating the VBO of the fully relaxed interface and then subtracting the electronic VBO of the atomically unrelaxed interface. For each system we have considered two different non-polar directions $\hat{{n}}_1$ and $\hat{{n}}_2$ and then the calculated $V_{IR}^{\hat{{n}}_1}$ and $V_{IR}^{\hat{{n}}_2}$ are compared in Fig. 6(a) (details in Table S3 in the SM). We see that the direction dependence associated with ionic relaxation is more obvious than that associated with electron relaxation, and this deviation increases for interfaces with larger $|V_{IR}^{\hat{{n}}}|$. The root-mean-square deviation of $V_{IR}^{\hat{{n}}}$ between $\hat{{n}}_1$ and $\hat{{n}}_2$  is 128 meV vs. 27 meV when considering only electron contributions. Thus, ionic relaxation worsens the orientation independence of band alignment as it inherintly breaks the local symmetry of the atoms. However, we note that when excluding the three outliers in Fig. 6(a) (MgS-SrO, AgCl-BaO and SrS-CsF), the root-mean-square deviation reduces to only 40 meV. Thus, excluding these outliers, both electronic and ionic contributions to the band alignment are orientation independent, which is directly in line with experiments. 

A screened ion model is developed to understand the formation of ionic relaxation dipoles wherein the ions are considered as point charges of nominal valence charge such that the ionic relaxation dipole is written as 
\begin{equation}
	\vec{{P}}_{IR} = \sum_{I} q_{I}({\vec{R}}^{'}_{I}-{\vec{R}}_{I}),
\end{equation}
where the sum is over displaced ions, ${\vec{R}}_{I}$ and ${\vec{R}}^{'}_{I}$ are the position of the $I$-th ion before and after ionic relaxation, and $q_{I}$ is the chemical valence. Applying Eq. (6) we obtain the screened ion model VBO shift,
\begin{equation}
	V_{IR, Model}^{\hat{{n}}} = \frac{\sigma_{IR}^{\hat{{n}}}}{\overline{\varepsilon}} = \frac{\vec{{P}}_{IR} \cdot \hat{{n}}}{\overline{\varepsilon}S_{\hat{{n}}}}, 
\end{equation}
where $\sigma_{IR}^{\hat{{n}}}=\vec{{P}}_{IR} \cdot \hat{{n}} / S_{\hat{{n}}}$ is the interfacial density of the ionic relaxation dipoles projected onto the $\hat{{n}}$ direction, $\overline{\varepsilon}$ is the averaged dielectric constant ($\overline{\varepsilon}=(\varepsilon_\infty^A+\varepsilon_\infty^B)/2$) of the two materials and $S_{\hat{{n}}}$ is same as Eq. (6) (details in Table S4 in the SM). The results of the screened ion model are presented in Fig. 6(b). Here we find that the model qualitatively agrees with the DFT results of the ionic relaxation VBO shift, however the values are systematically underestimated by a factor of approximately $1/3$, namely $V^{\hat{{n}}}_{IR,Model}/V^{\hat{{n}}}_{IR,DFT} \approx 2/3$ (Note S6 in the SM). Even so, we see that in Fig. 3(b) the same outliers have the largest deviation from the line. This indicates that differences in the magnitude of the ionic relaxation normal to the interface ($({\vec{R}}^{'}-{\vec{R}})\cdot \hat{{n}}$) for different  orientations are responsible for the emergence of direction dependence in the ionic relaxation contribution to VBO. Furthermore, as the magnitude of ionic relaxation increases, so too does the asymmetry in the VBO.

\section{Conclusion}

Using Wigner-Seitz atoms which obey the crystal symmetry to partition the bulk charge density, we examine the electron relaxation which occurs at the interface of a wide range of semiconducting heterojunctions. We show that this electron relaxation is localized to bonds at the interface and as a result the effect of interfacial electron relaxation can be faithfully modeled by point dipoles which exists along the bonds at the interface. Combining bulk contributions to the band offset, arising from the WSA quadrupole, and these interfacial bond dipoles we present the WSA+BD theory of band alignment which we show yield valence band offsets within 30 meV of those calculated using DFT. Constructing a geometric model of the interfacial dipole as originating from the sum of point bond dipoles, we show that the orientation dependence of the band alignment in isotropic and anisotropic systems, and for both polar and non-polar directions, can be understood in a unified framework. Not only does it correctly describe the orientation dependence of band alignment obtained via DFT for anisotropic materials, but the model shows that for isotropic systems the geometry of the system leads to strict orientation independece of the band alignment. In addition to providing a deeper understanding of band alignment, these results indicate that the preservation of atomic symmetry, as in the Wigner-Seitz atom partitioning, minimizes the perturbation manifested as local charge-transfer dipoles, suggesting the WSA partitioning may have deeper physical significance.

\textbf{Method:} DFT calculations are carried out by VASP code\cite{kresse1996} with PBE\cite{perdew1996} functional. The lattice structure of materials are taken from Material Porject\cite{jain2013} database. A plane-wave cutoff of 400 - 500 eV is used, depending on the elements involved in the interface system. The properties of Wigner-Seitz atoms/surfaces/interfaces are determined from bulk unit cell calculations. The atomistic DFT result of valence band offset is determined from a direct supercell calculation, in which the interface supercell is constructed by connecting the bulk supercells of two constituent materials along the interface normal direction. The interfacial electron density is fully relaxed with frozen atomic positions to determine the interfacial dipoles. Here the effect of interfacial atomic relaxation is ignored since only lattice matched materials are considered. The convergence of valence band offset on supercell length is examined to ensure an error $\le$ 0.01 eV.

\textbf{Supplementary Material:} See the supplementary material for more details, Notes S1–S6 and Tables S1-S4.

\textbf{Acknowledgments:} This work was supported by the U.S. DOE Grant No. DE-SC0002623. The supercomputer time sponsored by National Energy Research Scientific Center (NERSC) under DOE Contract No. DE-AC02-05CH11231 and the Center for Computational Innovations (CCI) at Rensselaer Polytechnic Institute (RPI) are also acknowledged.

\textbf{Conflict of Interest:} The authors have no conflicts to disclose.

\textbf{Ethics approval:} Neither animal nor human subjects are associated with this work.

\textbf{Author contributions:}  Z.Y.J. did the calculations. Z.Y.J., D.W., and S.B.Z. did the theoretical analyses. The paper is written by Z.Y.J. and D.W. with the help of S.B.Z., S.B.Z. proposed and initiated the project.

\textbf{Data availability:} Dataset that supports the findings of this study is available within the article and its supplementary material. Some calculations are performed with an in-house code, which will be available upon reasonable request.

\end{document}